# Extending Agents by Transmitting Protocols in Open Systems

Lavindra Priyalal de Silva
RMIT University,
Melbourne,
Australia
ldesilva@cs.rmit.edu.au

Michael Winikoff
RMIT University,
Melbourne,
Australia
winikoff@cs.rmit.edu.au

Wei Liu*
University of Western
Australia, Perth,
Australia
wei@csse.uwa.edu.au

## ABSTRACT
Agents in an open system communicate using interaction protocols. Suppose that we have a system of agents and that we want to add a new protocol that all (or some) agents should be able to understand. Clearly, modifying the source code for each agent implementation is not practical. A solution to this problem of upgrading an open system is to have a mechanism that allows agents to receive a description of an interaction protocol and use it. In this paper we propose a representation for protocols based on extending Petri nets. However, this is not enough: in an open system the source of a protocol may not be trusted and a protocol that is received may contain steps that are erroneous or that make confidential information public. We therefore also describe an analysis method that infers whether a protocol is safe. Finally, we give an execution model for extended Petri nets.

## 1. INTRODUCTION
An intelligent agent is situated, reactive, proactive, autonomous and social [9]. The social ability of an agent is exercised in a multi-agent system, which constitutes a collection of such agents. A multi-agent system where agents are created by different organisations using possibly different agent architectures is often termed an open system. For example, the Agentcities network.

Open systems introduce issues for agents that want to communicate. For example, issues of *privacy*, where an agent should ensure as much as possible that its privacy[1] is not jeopardised in communicating with another agent. *Safety* is another issue whereby an interaction should not require an agent to expend an unreasonable amount of resources for example.

Agents use interaction protocols to communicate with one another. If an agent's designer knows all the possible interaction protocols required for a new agent, the agent can be designed with the ability to communicate using this *fixed* set of protocols. This however stops agents in an open system from learning new protocols. The inability to learn also enforces a static open system that cannot be upgraded, for example by adding a new protocol.

We adopt the solution of agents communicating by transmitting interaction protocols. We fill in the gap in earlier research by providing a mechanism for transmitting protocols in the light of the above issues that open systems introduce. Firstly we propose a *representation* of interaction protocols to be transmitted, as extended Petri Nets[2]. Next we propose a method to *analyse* protocols received by another agent to check if they are adoptable (in terms of privacy, safety etc.), and finally an *execution* model for extended Petri Net protocols is proposed.

Consider the following example. A new agent $A$ is created for an open system. As a requirement, $A$ searches for online auctions that sell laptops and finds agent $B$ who is an auctioneer selling laptops. $A$ decides to communicate with $B$, and uses a generic 'request-reply'[3] protocol to request from $B$ a suitable protocol so that $A$ can participate (bid) in the auction. On this request, $B$ realises that $A$ is an agent that wants to enter the auction and sends $A$ a bidding protocol in a Petri Net representation. $A$ can now use the bidding protocol, but not before making sure it is *safe* to do so. For example, if the protocol requires $A$ to send a message to $B$ with $A$'s credit card number before bidding for an item, $A$ will decide not to use the protocol. If the protocol seems to be safe, $A$ will execute the bidding protocol which will result in it communicating with $B$. Due to space limitations it not possible to present the full details of this work in this paper. See [2] for more details.

### 1.1 Assumptions
In addition to the issues mentioned earlier, more basic but yet significant issues exist in open systems, such as *incompatibilities* that make it difficult for agents to communicate. Incompatibilities include different communication standards (such as FIPA or KQML), issues with understanding messages sent to one another[4], etc. To work around incompatibility issues, we assume that all agents adhere to a common communication standard and are therefore able to send messages to one another. Information received in a message is assumed to be understood by an agent using ontology information available to that agent. Much work is being done in the ontology area to support this. We also assume that all agents use the same "language" for encoding message content, such as XML.

### 1.2 Related Work
Although transmitting protocols have been proposed in earlier research [4, 5, 6], they were primarily for closed systems. Research such as [3] that had concentrated on open systems did not consider the issues of privacy, safety, etc. – significant when using a protocol sent by another agent. Similarly, research (for example [1]) that focused on descriptive protocols and protocol transmission left out issues arising as a result of this transmission, namely, issues of privacy and safety.

---

[0] Wei Liu was at RMIT at the time this research was carried out.
[1] An agent's need to keep certain information from being revealed to other agents.

[2] Refer to [8] for a detailed explanation of what Petri Nets are and how they work. Since Petri Nets are used to represent interaction protocols, we use the terms *Petri Net* and *Protocol* to mean the same thing.
[3] For our purposes, we assume that all agents in an open system can understand a generic 'request-reply' protocol which will be used to request for other protocols.
[4] For example, one agent could be using LISP syntax for representing content in messages and another could be using Prolog.

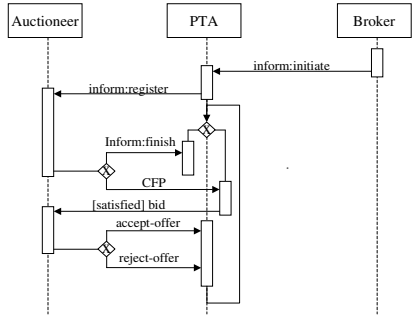

**Figure 1: AUML protocol for multiple participants**

## 2. REPRESENTING PROTOCOLS

In our research, protocols represent the interaction from a single agent's perspective, such as the *bidder* in the previous example. We call such protocols "local" protocols. Local protocols differ from "global" protocols, where the latter describes an interaction from a global perspective. A global protocol for the previous example include the interaction between the auctioneer, bidders and a possible supplier for example. Therefore according to our definition, AUML[5] represents protocols from a *global* perspective.

In addition to AUML, protocols have been represented using various notations [4, 5, 6] including Finite State Machines and Pushdown Automata, Dooley Graphs and Petri Nets. We have selected the Petri Net notation to represent local protocols because of its advantages over the other notations, such as the ability to represent concurrency in conversations. Refer to [7] for a comparison of these different protocol representations.

In addition to protocols being "local", we require protocols to be complete. A protocol should have enough detail so that an agent using it will know exactly what it needs to do at different stages in the interaction. This is in contrast to AUML for example, which does not capture internal processing. We have made protocols complete by enabling the specification of *actions* for an agent at different stages of an interaction. We have extended the basic Petri Net model to be able to represent external actions; messages to send and receive, and, internal actions; a specification of what functions to execute, what variables to read and write to and what conditions to test for.

An AUML protocol[6] is shown in Figure 1 along with the corresponding[7] extended Petri Net protocol for the Personal Travel Assistant (*PTA*) in the AUML protocol. The Petri Net extends the AUML protocol by adding internal processing details and defines the interaction from the *PTA* agent's perspective.

In the Petri Net, labels *Recv* and *Send* represent external actions,

---
[5] Agent Unified Modeling Language: A derivative of UML (Unified Modeling Language). http://www.auml.org/
[6] This AUML was created using the example at http://www.fipa.org/specs/fipa00080/XC00080B.html (section 5.3).
[7] The Petri Net represents our interpretation of the AUML protocol.

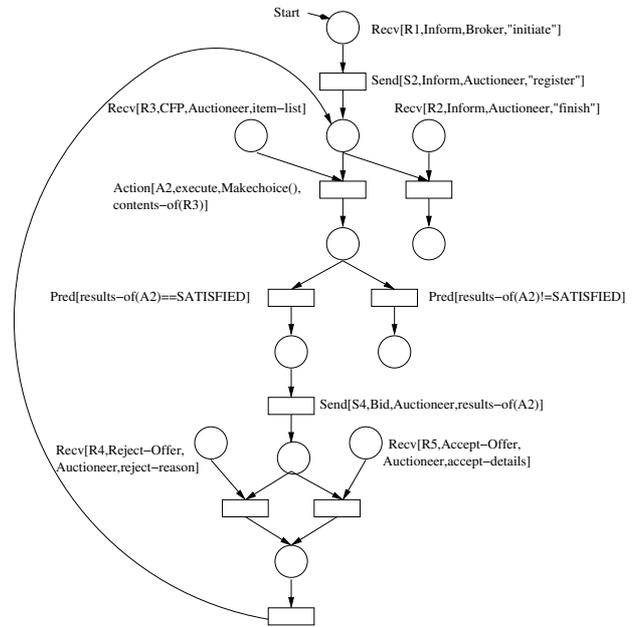

**Figure 2: Petri Net protocol for the *PTA* participant**

and *Action* and *Pred* represent internal actions. The *values* separated by commas next to each action specify arguments to that action. The prototype or legal syntax for how and what values are to be provided to actions is defined in the action's *template*, and is explained in sections that follow.

```
Send[<S-Label>,<Performative>,<Receiver>,<S-Content>]
Recv[<R-Label>,<Performative>,<Sender>,<R-Content>]
Action[<Act-Label>,<Type>,<Act>,<A-Content>]
Pred[<Boolean-Expression>]
```

### 2.1 Specifying Message Exchange

In our extended Petri Net notation, a specific message to be received by an agent executing the protocol is described by a *Recv* label. Recv labels are represented by Petri Net *places*. The agent from whom a message is to be received is specified in the *Sender* slot of the Recv. A Recv also requires other information such as the *R-Label* which uniquely identifies a Recv label on the Petri Net, the *Performative* of the message and the *content* of the message, which can be specified if known or ommitted otherwise. Similarly, a message to be sent by an agent will be specified in a Petri Net *transition* using a *Send* label. The Send requires similar information to that of the Recv. The *S-Content* in the Send can be used to specify a known content, or whether the message's content has to include results from a previously executed Action (see below). This is shown on the Petri Net in Send: *S4*, where $results\text{-}of(A2)$ refers to results from executing Action *A2*.

For example, the Petri Net in Figure 2 shows that the *PTA* begins by receiving a message from the *Broker* agent with an *Inform* performative (shown in the first Recv). It then sends a message to the auctioneer with a "register" note in its content by using an *Inform* performative.

### 2.2 Specifying Internal Functionality

The processing to be done involves function execution and reads and writes to variables. By allowing the specification of function executions and variable manipulations in protocols[8], ambiguity in protocols that may lead to different interpretations of protocols by different agents can be avoided.

The processing to be done is defined in Petri Net transitions using an *Action* label followed by its values. The type of action (i.e. *execute* for functions and *read* or *write* for variables) is specified in the *type* slot. The name of the function to be executed or the variable to be manipulated is specified in the *Act* slot. The argument(s) to the function or value for variables is specified in the *A-Content* slot. The *A-Content* slot can also be used to specify where (within the Petri Net) to obtain values from. For example, the Petri Net diagram shows how Action *A2* should obtain the argument(s) to the *Makechoice()* function when it is made available in the message received by the Recv labelled *R3*. The arguments to *Makechoice()* will be included in that message. Once *Makechoice()* is executed, its return value will be sent to the auctioneer in Send *S4* with a *Bid* performative as explained before.

We define predicates to be functions that return either True or False. Predicates are used to test conditions when the protocol is executed. Predicates are defined on *transitions* using *Pred* labels and prevent transitions from firing if false.

## 3. ANALYSING PROTOCOLS

Going back to the example mentioned in section 1, when agent *B* (the auctioneer) sends the bidding protocol to agent *A*, agent *A* will analyse it by firstly, seeing if it conforms to certain *syntax* and *semantic* requirements. Syntax checks are done to ensure that the protocol adheres to the proper extended Petri Net format. Examples include checking if the Petri Net begins with a place, ensuring that a value in an Action label's *Act* slot does not refer to a function that does not exist, etc. All resulting invalid *nodes*[9] after the syntax and semantic checks are marked "unsafe".

Next the Petri Net is checked for *loops*[10] that lead to problems when the protocol is executed. Finally, a check is made to ensure that the protocol sent by *B* is not a threat to *A*'s *privacy*.

### 3.1 Protecting Private Information

The privacy of an agent is jeopardised if it has to release private information by executing the protocol. By private information we mean information belonging to an agent (or its user(s)) that the agent does not want to reveal to all other agents, such as credit card information, driver's license etc.

An invasion of privacy becomes an issue when executing a received protocol if the protocol is created by a remote, untrustworthy agent. An agent that creates a protocol can, via Action transitions, potentially execute *any*[11] function, or, read or modify *any* variable in the agent who executes that protocol. An agent could therefore execute a protocol that defines an Action and Send that would make the agent read and send its driver's license number to another agent for example.

One way of protecting privacy is for each agent to maintain an access list, with information such as; which other agents are allowed access to which variables and functions and what kind of access is permitted (read, write execute). Action transitions can be analysed with this information and unsafe Actions can be detected.

However, assigning fixed access permissions to functions and variables is not always feasible. Some functions and variables may need to be allowed access at certain times but denied access at other times. For example, a variable representing an agent's credit card number may need to be kept private by that agent until an item that the agent buys with that credit card is delivered. More precisely, conditions to allow an agent's credit card number to be read (this read is identical to paying for an item and can be called *Pay*) by another agent are that, firstly, the protocol should specify that an agent should want to buy an item before the item is paid for (call this condition *WantToBuyItem*). Secondly the protocol should indicate that at some time in the future, an item paid for will indeed be delivered (call this condition *Delivery*).

*WantToBuyItem* and *Delivery* are *preconditions* to *Pay*. A precondition is a condition that needs to hold before an action (or number of actions) can be executed. For example, before paying for an item we should want to buy the item. We identify two types of preconditions: past and future. A *past* precondition is where the condition is expected to hold at the time that the action is to be performed. For example, wanting to buy the item before paying for it. A *future* precondition is where the condition is not expected to hold at the time that the action is to be performed. For example, paying for an item before the item is received. In this case receiving the item is a precondition since we do not want to pay unless we know that the item will be delivered. However, at the time payment is made, the item will not yet have been received. This issue is considered in section 4.

We use *Action Templates* to define the past and future preconditions for actions. Action Templates can be used with Petri Net actions to ensure that the latter adheres to the necessary past and/or future precondition(s). The Action Template for the previous example is shown below. *Pay* represents the actions of paying for an item, outgoing arcs are future preconditions and incoming arcs are past preconditions. *Delivery* can either be a past or future precondition and *WantToBuyItem* is the past precondition.

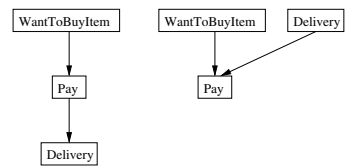

In our research, we have provided a mechanism to map an Action Template to a specific set of actions in a protocol to check if the latter are safe. A set of actions is unsafe if a past (or future) precondition is defined for it on an Action Template which is not reflected on the on the Petri Net. For further details, refer to [2].

### 3.2 Propagating Safety Information

At the completion of the stages of protocol analysis, the Petri Net protocol may have some places and/or transitions marked "unsafe". Those nodes should be avoided when the Petri Net is executed. If even one of those unsafe nodes cannot be avoided, the Petri Net protocol simply cannot be used.

Inferring whether an unsafe node is avoidable or not is done by a method we call *Safety Propagation*. In this method, given an unsafe node, we "propagate" from that node to find other nodes that are unsafe or *unusable*. During propagation, a node is marked unsafe if its execution will result in the execution of the unsafe node being considered. A node is marked unusable if it cannot be reached as a result of a node that has been marked unsafe or unusable. If the entire Petri Net or its initial place is unsafe or unusable, or if all of the final places are unsafe or unusable, the Petri Net will be rejected.

---

[8]The *functions* and *variables* of an agent will be defined in an ontology. Agents can therefore use this ontology information to exchange protocols with other agents that understand them.
[9]Places and transitions
[10]Omitted due to space constraints. Refer to [2] for details.
[11]Within the constraints of the ontology

## 4. EXECUTING PROTOCOLS

If the protocol is not rejected, it will be executed. The execution of a Petri Net is the movement of tokens from a place to another connected place, as a result of transitions firing. Since our extended Petri Net requires internal and external actions to be performed during its execution, to deal with these, we have made minor modifications to the Petri Net execution model. Since some aspects of Petri Net execution were covered in section 2, the next paragraph serves as a summary.

During execution, a place that has a Recv label gets one token placed on it whenever the appropriate message (as specified by values for that Recv) is received by the agent. A transition that has a Send label sends the message described by its values whenever that transition fires, which may also involve obtaining results from a previously executed Action. A transition that has a Pred label requires a change to the way that transitions usually fire. We require the predicate returning True to be an additional criteria for that transition to fire[12]. A transition that has an Action executes the internal action defined by its values before that transition fires and will fire once the internal action has completed execution.

Additionally, the safety of a protocol is a characteristic that can sometimes only be tested during execution time. For example, an action cannot be considered safe if its future precondition is based on a runtime factor such as the receipt of a message from another agent, because, the action could be executed but the message required to satisfy its future precondition may never be received. This reinstates the privacy issue discussed earlier since the agent may release private information (in paying for an item for example) but may not be sent the item afterwards. We provide a partial solution to this by requiring that agents request a *guarantee* from the remote agent(s) to send the message(s) required to satisfy a future precondition, immediately before the action (e.g. pay) to that future precondition is executed[13]. If the future precondition is guaranteed, the requesting agent will in turn execute the action. This is similar to saying, "If I send you my Credit Card Number now, will you promise to send me the Receipt (for the item bought) afterwards". No such guarantee is required for past preconditions since a past precondition to an action will have already been executed when the action is being considered.

We also provide a mechanism for the remote agent to infer whether it can guarantee some other agent's future precondition. The remote agent does this by mapping a guarantee request received against its own protocol to test if it can indeed send the message(s) requested. If a guarantee cannot be sent, the protocol will be terminated (after negotiation) by the agent requiring the guarantee.

---

[12] In addition to all incoming places requiring at least one token on them.

[13] This is only a partial solution because an agent may guarantee the sending of a message and may not send the message afterwards. Research based on *Trust Networks* deals with such issues.

## 5. FUTURE WORK AND CONCLUSION

Due to time constraints, our work was mostly theoretical. The completion and integration of all the algorithms proposed into a system that can transmit, analyse and execute protocols is left as future work. Furthermore, the details of how a protocol that is relevant to a given situation is chosen (see example in section 1), and a more precise description of Action Templates is also left as future work. In our research, we have presented a series of detailed requirements for agents, to be able to transmit protocols in open systems. We have considered issues such as privacy and safety that become significant when a protocol received from an untrustworthy agent is to be used. In order to use protocols, we have proposed a Petri Net representation for them that can be analysed to deduce their safety, and provided agents with an option to reject a protocol if it were not safe. If the safety could not be confirmed during analysis time, the agent is able to terminate a protocol during its execution if a required guarantee(s) was not met.

## 6. ACKNOWLEDGEMENTS

Thanks to Lin Padgham, James Harland and the RMIT Agent Group for useful feedback on the Honours Thesis. Thanks to the FIPA mailing list for helping us find relevant information. We would also like to acknowledge the support of the Australian Research Council (Linkage grant LP0218928).